\documentstyle[preprint,aps,prb,epsfig]{revtex}
\begin{document}
\draft
\pagestyle{plain}
\newcommand{\D}{\displaystyle}
\newcommand{\ab}{\frac{a}{b}}
\newcommand{\TT}{\frac{3}{2}}
\title{\bf The Critical Temperature of an Anisotropic Superconductor
in the Presence of a Homogeneous Magnetic Field and Impurities}
\author{G. Hara\'n\cite{AA}, J. Taylor and A. D. S. Nagi}
\vspace{0.4cm}
\address{Department of Physics,
University of Waterloo,
Waterloo, Ontario,
Canada, N2L 3G1}
\maketitle

\begin{abstract}
The effect of a homogeneous magnetic field and nonmagnetic 
impurities on the critical temperature of an anisotropic 
superconductor has been investigated. The role of these pair-
breakers in relation to the anisotropy of the order parameter 
is clarified.\\ 


\vspace{1cm}
\noindent
{\it Reprinted from Solid State Communications, Vol. 100, No. 3,\\ 
G. Hara\'n, J. Taylor and A. D. S. Nagi,\\ 
The Critical Temperature of an Anisotropic Superconductor
in the Presence of a Homogeneous Magnetic Field and Impurities, pp. 173-175, 
1996,\\ with permission from Elsevier Science Ltd, The Boulevard, 
Langford Lane, Kidlington 0X5 1GB, UK} 
\end{abstract}
\newpage

There now exists a considerable experimental evidence supporting 
the d-wave superconductivity in the cuprates, reviewed recently  
by Annett et al. 
\cite{1} The nonmagnetic impurities are known to destroy the 
Cooper pairs in anisotropic superconductors. Also a homogeneous 
magnetic field is a pair breaker in a singlet state as it separates 
the up and down spin bands. It is important to study the role 
of above two pair breakers in relation to the anisotropy of the 
order parameter. Such an investigation has been carried out in the 
present paper. 
Our results would apply to 
an impure thin superconducting film of a thickness less than the 
London penetration depth in a parallel magnetic field.\\

Consider an anisotropic superconductor in the presence of a magnetic  
field $H$ parallel to the ab-plane and having a random 
distribution of nonmagnetic impurities. Treating the electron-impurity 
scattering within first-order Born approximation and neglecting the 
impurity-impurity interaction, \cite{2} the normal 
$\hat{G}\left(\tilde{\omega},{\bf k}\right)$ and the anomalous 
$\hat{F}\left(\tilde{\omega},{\bf k}\right)$ Green's functions 
for the superconducting electrons averaged over the impurity positions 
are given by ($z$ axis of the spin space is chosen along the magnetic 
field and $\hbar=k_{B}=1$) 

\begin{equation}
\label{e1}
\hat{G}\left(\tilde{\omega},{\bf k}\right)=\left(
\begin{array}{cc}
G_{+}\left(\tilde{\omega},{\bf k}\right) & 0\\
0 & G_{-}\left(\tilde{\omega},{\bf k}\right)
\end{array}
\right)
\end{equation}

\begin{equation}
\label{e2}
\hat{F}\left(\tilde{\omega},{\bf k}\right)=\left(
\begin{array}{cc}
0 & F_{-}\left(\tilde{\omega},{\bf k}\right)\\
-F_{+}\left(\tilde{\omega},{\bf k}\right) & 0
\end{array}
\right)
\end{equation}

\noindent
with

\begin{eqnarray}
\label{e5}
G_{\pm}\left(\tilde{\omega},{\bf k}\right) & = &
-\frac{i\tilde{\omega}_{\pm}+\xi_{k}}
{\tilde{\omega}_{\pm}^{2}+\xi_{k}^{2}+|\tilde{\Delta}_{\pm}\left(
{\bf k}\right)|^{2}}
\end{eqnarray}

\begin{eqnarray}
\label{e5a}
F_{\pm}\left(\tilde{\omega},{\bf k}\right) & = &
\frac{\tilde{\Delta}_{\pm}\left({\bf k}\right)}
{\tilde{\omega}_{\pm}^{2}+\xi_{k}^{2}+|\tilde{\Delta}_{\pm}\left(
{\bf k}\right)|^{2}}
\end{eqnarray}

\noindent
where the subscripts $+(-)$ apply to a spin parallel (antiparallel) 
to the z-axis. Further, $\xi_{k}$ is the quasiparticle energy, 
${\bf k}$ is wave vector. The renormalized Matsubara frequency 
$\tilde{\omega}_{\pm}$ and the renormalized order parameter 
$\tilde{\Delta}_{\pm}\left({\bf k}\right)$ are given by 

\begin{eqnarray}
\label{e6}
\tilde{\omega}_{\pm} & = & \omega_{\pm} +
in_{i}\int |v\left({\bf k}-{\bf k'}\right)|^{2}
G_{\pm}\left(\tilde{\omega},{\bf k'}\right)\frac{d^{3}k'}
{\left(2\pi\right)^{3}}
\end{eqnarray}

\begin{eqnarray}
\label{e6a}
\tilde{\Delta}_{\pm}\left({\bf k}\right) & = & \Delta\left({\bf k}\right)
+n_{i}\int |v\left({\bf k}-{\bf k'}\right)|^{2}
F_{\pm}\left(\tilde{\omega},{\bf k'}\right)\frac{d^{3}k'}
{\left(2\pi\right)^{3}}
\end{eqnarray}

\noindent
In above $\omega_{\pm}=\omega\pm ih$ ($h=\mu_{B}H$, $\mu_{B}$ is the 
Bohr magneton), $\omega=\pi T(2n+1)$ (T is temperature and n is an 
integer), 
$\Delta\left({\bf k}\right)$ is superconducting order parameter, 
$n_{i}$ is the impurity concentration and 
$v\left({\bf k-k'}\right)$ is the impurity potential.\\

We consider a singlet order parameter with its orbital part given by 

\begin{equation}
\label{e4}
\Delta\!\left({\bf k}\right)=\Delta e\!\left({\bf k}\right)
\end{equation}

\noindent
where $e\left({\bf k}\right)$ is a basis function of a one
dimensional (1D) irreducible representation of an appropriate symmetry point
group, which seems to be a good approximation for high $T_{c}$ superconductors 
.\cite{1}  We normalize $e\!\left({\bf k}\right)$ 
by taking $\left<e^{2}\right>=1$,
where $<...>$ denotes the average value over the Fermi surface.
We may mention that $0\le \left<e\right>^{2} \le 1$. 
$\left<e\right>^{2}=1$ corresponds to an isotropic superconductor 
and $\left<e\right>^{2}=0$ seems to be a good approximation for 
a d-wave $x^{2}-y^{2}$ state in YBCO.\\

To proceed further, we replace $\int d^{3}k/\left(2\pi\right)^{3}$
by $N_{0}\int_{FS}dS_{k}n\left({\bf k}\right)\int d\xi_{k}$, 
where $N_{0}$ is
the overall electron density of states per spin at the Fermi surface (FS),
$n\left({\bf k}\right)$ is the angle resolved FS density
of states and $\int_{FS}dS_{k}$ denotes integration over the Fermi surface;
$n\left({\bf k}\right)$ is normalized to unity, i.e. $\left<n\right>=1$.\\
\noindent
Using Eqs. (\ref{e5}) and (\ref{e5a}) in Eqs. (\ref{e6}) and (\ref{e6a}), 
taking  $v\left({\bf k-k'}\right)$ as   
independent of momentum and performing the $\xi_{k}$ integration
(particle-hole symmetry of quasiparticle spectrum is assumed)
we find, 

\begin{eqnarray}
\label{e7}
\tilde{\omega}_{\pm}&=&\omega_{\pm}\left(1
+u_{\pm}\left(\omega\right)\right)\\
\tilde{\Delta}_{\pm}\left({\bf k}\right)&=&
\Delta\left(e\left({\bf k}\right)
+e_{\pm}\left(\omega\right)\right)
\end{eqnarray}

\noindent
where $u_{\pm}\left(\omega\right)$ and $e_{\pm}\left(\omega\right)$ functions
are determined by the self-consistent equations which read

\begin{eqnarray}
\label{e8}
u_{\pm}\left(\omega\right)&=&\Gamma\int_{FS}dS_{k}n\left({\bf k}\right)
\frac{1+u_{\pm}\left(\omega\right)}{\left[\tilde{\omega}_{\pm}^{2}+
|\tilde{\Delta}_{\pm}\left({\bf k}\right)|^{2}\right]^{1/2}}
\end{eqnarray}

\begin{eqnarray}
\label{e8a}
e_{\pm}\left(\omega\right)&=&\Gamma\int_{FS}dS_{k}n\left({\bf k}\right)
\frac{e\left({\bf k}\right)+e_{\pm}\left(\omega\right)}
{\left[\tilde{\omega}_{\pm}^{2}+
|\tilde{\Delta}_{\pm}\left({\bf k}\right)|^{2}\right]^{1/2}}
\end{eqnarray}

\noindent
with $\Gamma=\pi N_{0}n_{i}|v|^{2}$.\\

The self-consistent equation for the matrix order parameter  
$\hat{\Delta}\left({\bf k}\right)=\Delta\left({\bf k}\right)i\sigma_{y}$ 
($\sigma_{y}$ is Pauli matrix) reads

\begin{equation}
\label{e9}
\D\hat{\Delta}_{\alpha\beta}\left({\bf k}\right)=
-\frac{T}{2}\sum_{\omega}\sum_{\gamma,\nu}  
\int \frac{d^{3}k'}{\left(2\pi\right)^{3}}
V\left({\bf k},{\bf k'}\right)
\left(\delta_{\alpha\gamma}\delta_{\beta\nu}
-\delta_{\alpha\nu}\delta_{\beta\gamma}\right)
\hat{F}_{\gamma\nu}\left(\omega,{\bf k'}\right)
\end{equation}

\noindent
where $V\left({\bf k}, {\bf k'}\right)$ is the phenomenological pair 
potential taken as

\begin{equation}
\label{e10}
V\left({\bf k}, {\bf k'}\right)=-V_{0}e\left({\bf k}\right)
e\left({\bf k'}\right)
\end{equation}

\noindent
Using Eqs. (\ref{e2}), (\ref{e4}) and (\ref{e10}) with Eq. (\ref{e9}),  
we obtain

\begin{equation}
\label{e11}
\Delta=\frac{1}{2}N_{0}V_{0}
T\sum_{\omega}\int_{FS}dS_{k}n\left({\bf k}\right)
e\left({\bf k}\right)\int d\xi_{k} \left(
F_{+}\left(\omega,{\bf k}\right)+F_{-}\left(\omega,{\bf k}\right)\right)
\end{equation}

\noindent
and after integration over the quasiparticle energy we get

\begin{equation}
\label{e12}
\begin{array}{l}
\Delta=\D \frac{1}{2}N_{0}V_{0}
\pi T\sum_{\omega}\int_{FS}dS_{k}n\left({\bf k}\right)
e\left({\bf k}\right)\times\\
\\
\D\times\left(\frac{\tilde{\Delta}_{+}\left({\bf k}\right)}
{\left[\tilde{\omega}_{+}^{2}+|\tilde{\Delta}_{+}\left(
{\bf k}\right)|^{2}\right]^{\frac{1}{2}}}+
\frac{\tilde{\Delta}_{-}\left({\bf k}\right)}
{\left[\tilde{\omega}_{-}^{2}+|\tilde{\Delta}_{-}\left(
{\bf k}\right)|^{2}\right]^{\frac{1}{2}}}\right)
\end{array}
\end{equation}      

\noindent
Following standard procedure, \cite{5,6} we obtain the  
equation for the critical temperature $T_{c}$ as 

\begin{equation}
\label{e15}
\D\ln\left(\frac{T_{c}}{T_{c_{0}}}\right)=
\pi T\sum_{\omega}\left(f_{\omega_{\Delta=0}}
-\frac{1}{|\omega|}\right)
\end{equation}

\noindent
where $T_{c_{0}}$ is the critical temperature in the absence of 
magnetic field and impurities and 

\begin{equation}
\label{e13}
f_{\omega}=
\D\frac{1}{2}\int_{FS}dS_{k}n\left({\bf k}\right)
e\left({\bf k}\right)
\left(\frac{e\left({\bf k}\right)+e_{+}\left(\omega\right)}
{\left[\tilde{\omega}_{+}^{2}+|\tilde{\Delta}_{+}\left(
{\bf k}\right)|^{2}\right]^{\frac{1}{2}}}+
\frac{e\left({\bf k}\right)+e_{-}\left(\omega\right)}
{\left[\tilde{\omega}_{-}^{2}+|\tilde{\Delta}_{-}\left(
{\bf k}\right)|^{2}\right]^{\frac{1}{2}}}\right)
\end{equation}

\noindent
In order to obtain $f_{\omega_{\Delta=0}}$, we need values of 
$u_{\pm}\left(\omega\right)$ and $e_{\pm}\left(\omega\right)$ 
in $\Delta=0$ limit which are easily obtained from Eqs. (\ref{e8}) 
- (\ref{e8a}) and read 

\begin{equation}
\label{e17}
{u_{\pm}}^{0}\left(\omega\right)=
\frac{\Gamma}{\omega_{\pm}}sign\left(\omega\right)
\end{equation}

\begin{equation}
\label{e18}
{e_{\pm}}^{0}\left(\omega\right)=
\left<e\right>\frac{\Gamma}{\omega_{\pm}}sign\left(\omega\right)
\end{equation}

\noindent
Substituting ${\tilde{\omega}_{\pm}}^{0}=\omega_{\pm}\left(1
+{u_{\pm}}^{0}\left(\omega\right)\right)$ and 
${e_{\pm}}^{0}\left(\omega\right)$ into Eq. (\ref{e13}) and performing 
the integration over FS, $f_{\omega_{\Delta=0}}$ is easily obtained. 
Then Eq. (\ref{e15}) gives 

\begin{equation}
\label{e19}
\begin{array}{l}
\D\ln\left(\frac{T_{c}}{T_{c_{0}}}\right)=
\left(\left<e\right>^{2}-1\right)Re
\left(\psi\left(\frac{1}{2}+\frac{\Gamma}{2\pi T_{c}}
+\frac{ih}{2\pi T_{c}}\right)
-\psi\left(\frac{1}{2}\right)\right)\\
\\
\D -\left<e\right>^{2}Re
\left(\psi\left(\frac{1}{2}+\frac{ih}{2\pi T_{c}}\right)
-\psi\left(\frac{1}{2}\right)\right)
\end{array}
\end{equation}

\noindent
where $\psi\left(z\right)$ is the digamma function. \cite{7}\\

Eq. (\ref{e19}) clarifies the role of the two pair-breakers (
impurities and the magnetic field) in relation to the anisotropy 
of the order parameter. For $\left<e\right>^{2}=1$ (isotropic 
superconductor), Eq. (\ref{e19}) gives    

\begin{equation}
\label{e20}
\D\ln\left(\frac{T_{c}}{T_{c_{0}}}\right)=
Re\left(\psi\left(\frac{1}{2}\right)-
\psi\left(\frac{1}{2}+\frac{ih}{2\pi T_{c}}\right)\right)
\end{equation}

\noindent
which is a standard result. \cite{8,3} Here $T_{c}$ does not depend  
on the nonmagnetic impurities in agreement with Anderson's 
theorem. \cite{4} For $\left<e\right>^{2}=0$, Eq. (\ref{e19}) yields 

\begin{equation}
\label{e21}
\D\ln\left(\frac{T_{c}}{T_{c_{0}}}\right)=
Re\left(\psi\left(\frac{1}{2}\right)-
\psi\left(\frac{1}{2}+\frac{\Gamma}{2\pi T_{c}}
+\frac{ih}{2\pi T_{c}}\right)\right)
\end{equation}

\noindent
which gives the maximum pair-breaking effect of the impurities. Now 
the influence of the two pair-breakers is additive. For in between 
values of $\left<e\right>^{2}$ both terms on the right of  Eq. (\ref{e19}) 
contribute and the impurity pair-breaking  effect is in between that 
given by Eqs. (\ref{e20}) and (\ref{e21}).\\

Our results for the dependence of $T_{c}/T_{c_{0}}$ on 
normalized magnetic field $\kappa h/\pi T_{c_{0}}$ for different values 
of normalized scattering rate $2\kappa\Gamma/\pi T_{c_{0}}$, where 
$\kappa=e^{\gamma}\simeq 1.781$ and $\gamma$ is the Euler's constant,  
are shown in Fig. 1. We have taken $\left<e\right>^{2}=0$.  
From these results we make the 
following remarks: (1) In various curves, region of second (first) 
order phase transition is shown by solid 
(dashed) curve. The second and the first order phase transition  
curves meet at the tricritical point $t_{0}$. 
At that point Eq. (\ref{e15}) is satisfied and also the function 
$f_{1}=-\left[\left(2\pi T\right)^{3}/2\right]\sum_{\omega}
\left[df_{\omega}/d\Delta^{2}\right]_{\Delta=0}$ becomes zero. \cite{9} 
The coordinates of $t_{0}$ for $2\kappa\Gamma/\pi T_{c_{0}}$= 0, 0.2, 0.4  
are (0.608, 0.561), (0.549, 0.429) and (0.483, 0.280), respectively. 
It may be noted that the tricritical point appears at lower magnetic 
field as scattering rate is increased; (2) For $\Gamma=0$ (i.e. 
no impurities), the curve is the same as for an isotropic superconductor;  
(3) For a fixed value of the scattering rate, 
$T_{c}$ decreases with the increase of the magnetic field; 
(4) For a fixed value of $h$, $T_{c}$ decreases with the increase of the 
scattering rate.\\

In summary, we have studied the role of the  
pair breaking effects of the 
impurities and a homogeneous magnetic field in relation to  
anisotropy of the superconducting order parameter. 
Experiments on thin cuprate superconducting films 
(thickness less than the London penetration depth) in the presence 
of impurities and a parallel magnetic field are desirable to verify our 
results. Knowing the function $f_{1}$,  
the tricritical point  
and the jump in specific heat at $T_{c}$ can be calculated. These detailed 
calculations are in progress.\\

\noindent
Acknowledgement - This work is supported in part by the Natural Sciences 
and Engineering Research Council of Canada.

\newpage
\begin{center}
\begin{figure}[p]
\parbox{0.75cm}{\vfill $$ \frac{T_c}{T_{c_{0}}} $$ \vfill }
\parbox{6.5cm}{\epsfig{file=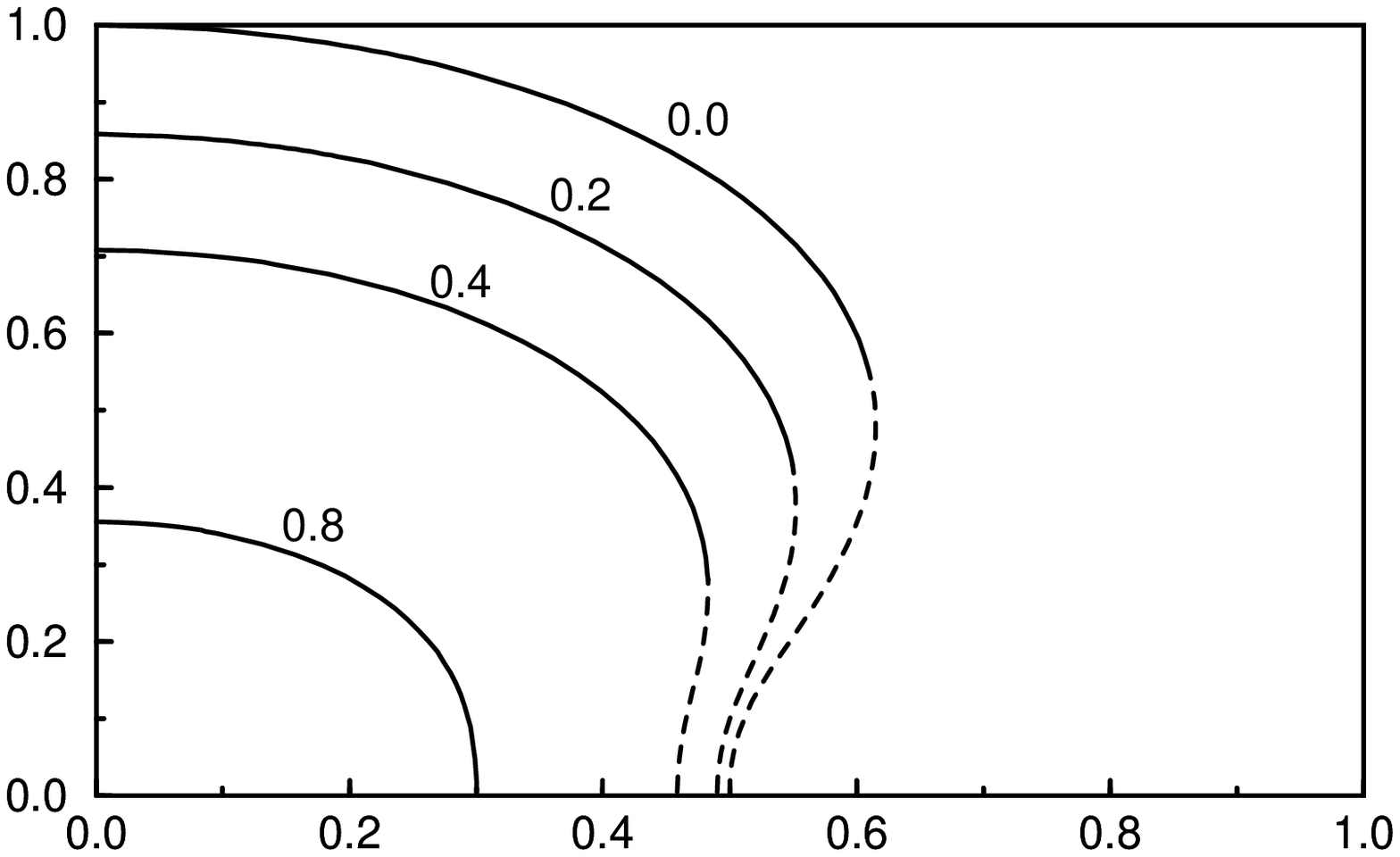,height=6.5cm,width=6.5cm} }
\parbox{1.8cm}{\hfill}
\parbox{8cm}{$$ \frac{\kappa h}{\pi T_{c_{0}}} $$}
\caption{
The normalized critical temperature $T_{c}/T_{c_{0}}$ 
(solid curve) as a function  
of the normalized magnetic field $\kappa h/\pi T_{c_{0}}$  
for the normalized impurity scattering rates $2\kappa\Gamma/\pi T_{c_{0}}
=0.0, 0.2, 0.4, 0.8$.  
The dashed curve shows the region of the first order 
phase transition. In all cases $\left<e\right>^{2}=0$ is taken.}
\end{figure}
\end{center}

\end{document}